\documentclass{elsart_n}
\journal{Acta Phys. Polon. B, Vol. 40, No. 9, 2519}
\date{Sept 2009}
\usepackage{epsfig}
\usepackage{fontenc}
\usepackage{axodraw4j}
\usepackage{pstricks}
\usepackage{color}
\usepackage[dvips, bookmarks, bookmarksnumbered, bookmarksopen,colorlinks=true,linkcolor=black]{hyperref}
\usepackage{ccaption}

\begin{document}

\newcommand{\nucl}[2]{{}^{#1}\mbox{#2}}

\begin{frontmatter}
\title{Study of Pion Production in $\nu_{\mu}$ CC Interactions on
$\nucl{16}{O}$ Using Different MC Generators}
{\large Ladek Zdr\'{o}j students Monte Carlo project}
{\normalsize 
\author[LNGS]        {Maddalena~Antonello}, 
\author[LAquila]        {Vincenzo~Caracciolo}, 
\author[Liverpool]        {Georgios~Christodoulou}, 
\ead                          {georgios@hep.ph.liv.ac.uk}
\author[London]        {James~Dobson}, 
\ead                          {j.dobson07@imperial.ac.uk}
\author[Bern]        {Eike~Frank}, 
\author[Wroclaw]        {Tomasz Golan}, 
\author[Dubna]        {Viacheslav~Lee},
\ead                          {lee@nusun.jinr.ru}
\author[Katowice]        {Slawomir~Mania},
\author[SoltanNucl]        {Pawel~Przewlocki},
\author[Bern]        {Biagio~Rossi},
\author[Cracow]        {Dorota~Stefan},
\ead                          {Dorota.Stefan@ifj.edu.pl}
\author[SoltanNucl,WarsawTech]        {Robert~Sulej}, 
\author[Katowice]        {Tomasz~Szeglowski},
\author[Regina]        {Roman~Tacik},
\author[Cracow]        {Tomasz~Wachala}
}
\address[LNGS]          {INFN-Laboratori Nazionali del Gran Sasso, Assergi, Italy}
\address[LAquila]          {University of L'Aquila, Italy}
\address[Liverpool]          {Liverpool University, UK}
\address[London]          {Imperial College London, UK}
\address[Bern]          {LHEP, University of Bern, Switzerland}
\address[Wroclaw]          {University of Wroclaw, Poland}
\address[Dubna]		{Joint Institute for Nuclear Research, Dubna, Russia}
\address[Katowice]          {University of Silesia, Katowice, Poland}
\address[Cracow]          {The Henryk Niewodniczanski Institute of Nuclear Physics PAN, Cracow, Poland}
\address[SoltanNucl]	{Soltan Institute for Nuclear Studies, Warsaw, Poland}
\address[WarsawTech]	{Warsaw University of Technology, Poland}
\address[Regina]          {University of Regina, Canada}
\begin{abstract}
{\small
In this report we present simulated event numbers, for various MC generators, for pion production in $\nu_{\mu}$ CC reactions on $\nucl{16}{O}$. For the simulation we used four different neutrino interaction generators: GENIE, FLUKA, NEUT, and NuWro, as proposed during the 45th Karpacz Winter School on neutrino interactions \cite{Ladekschoolweb}. 
First we give a brief outline of the theoretical models relevant to pion production. We then present results, in the form of tables showing the occupancy of primary and final state pion topologies, for all the generated samples. Finally we compare the results from the different generators and draw conclusions about the similarities and differences. For some of the generators we explore the effect of varying the axial mass parameter or the use of a different nuclear model.
}
\end{abstract}

\end{frontmatter}


\tableofcontents

\section{Introduction}
\label{sec:intro}
Understanding Charged-Current (CC) neutrino--nucleus interactions in the few GeV region is very important for many current and future neutrino experiments. The study of neutrino--nucleus reactions in this region is complicated and requires many intermediate steps, such as a description of the nuclear model, understanding the neutrino--nucleon cross sections, modeling of hadronization, as well as the modeling of intranuclear hadron transport and other secondary interactions. These can all play a significant role in how we understand the nature of neutrinos as well as providing useful information about nuclear phenomena. Because of this there are a number of Monte Carlo generators and numerical packages dedicated to the description of neutrino interactions: GENIE~\cite{Genie}, GiBUU~\cite{GiBUU}, FLUKA~\cite{FLUKA}, NEUT~\cite{NEUT}, NuWro~\cite{NuWro} and Nuance~\cite{Casper:2002sd} represent a large fraction of such generators and were all presented at the Ladek Winter School~\cite{Ladekschoolweb}.

During the school we undertook a project to compare the predictions made by different generators. Of the generators mentioned previously we looked at GENIE, NEUT, Nuance\footnote{The results of the simulation using Nuance alongside a description of the generator are presented in the appendix, Table~\ref{table:nuance}.}, FLUKA and NuWro. For now GiBUU is not included in this study because it was difficult to find a consistent way of comparing it to the other simulation packages\footnote{GiBUU uses a more sophisticated model in which the propagation/collision of resonances is handled explicitly. In this model the primary hadronic system is the QE nucleon, the resonances, and any pions from the non-resonant background. And so comparison with the other simulation packages is difficult as they decay the resonances before rescattering and so define the initial state as the decay products of all resonances.}. It was decided that we would focus our investigations on the production of pions in neutrino-nucleus interactions\footnote{The production of other mesons, like $\eta$ or $\rho$, is also possible but in general pion production dominates.}. This was because they form an important background in many neutrino oscillation experiments and are also theoretically challenging, due to processes such as final state interactions (FSIs).  A special web site of ``Ladek MC Project'' was set up to collate and discuss the results~\cite{Ladekprojectweb}. 

For each generator we produced a sample of mono-energetic (1 GeV) $\nu_{\mu}$ interactions on~$\nucl{16}{O}$, using the default settings for each generator. To simplify the analysis we consider only charged current (CC) interactions. Each generator had the following processes enabled: quasi-elastic~(QE) scattering, resonance~(RES) production, deep-inelastic scattering~(DIS) and coherent~(COH) pion production. We then analyzed the samples for each generator by looking at the various pion topologies before and after any secondary interactions.

Before presenting the results of the study we give a brief outline of the theoretical models of relevance to CC neutrino induced pion production.

\section{Overview of Neutrino-induced Pion Production}
\label{sec:theory}

As mentioned previously, the modeling of neutrino-nucleus interactions is complex and requires linking together many different pieces of theory. Here we focus on neutrino-nucleon cross-sections and how they are embedded in a nuclear environment. We leave the description of the hadronization and final state interaction models, often specific to a particular generator, to chapter~\ref{sec:numbers}.    

The total cross section for neutrino--nucleon scattering has the following form~\cite{Kuzmin:2006dt}
\begin{equation}
\sigma_{\nu N}^{ tot } = \sigma_{\nu N}^{ (Q)ES } \oplus \sigma_{\nu N}^{ 1 \pi } \oplus \sigma_{\nu N}^{ 2 \pi } \oplus \dots \oplus \sigma_{\nu N}^{ 1K } \oplus \dots \oplus \sigma_{\nu N}^{ DIS } .
\end{equation}
The region of neutrino energies around 1 GeV is particularly troublesome. It is in this region that many of the above cross sections are similar in magnitude. Here resonance single pion production contributes $\sim30 \%$ to the total cross section~\cite{Costas_Ladekschoolweb}, similar to the contributions from QEL and DIS processes. This is a problem experimentally as RES events can have indistinguishable signatures to DIS events in a detector, making it hard to measure each process exclusively.

Charged current QEL scattering of a neutrino on a free nucleon ($\nu_\ell\! +\! N \!\to\! \ell \!+\! N'$) is usually described using the Llewellyn Smith formalism~\cite{Llewellyn Smith:1971zm}. Although no pions are produced directly it is possible, through FSIs, to produce them in the final state system. Inside the nucleus, hadrons can be scattered elastically or inelastically, can be absorbed or charge exchanged and even produce extra pions (pion production). Thus, a small number of events with pions in the final state are expected from CCQE events, even though no pions were produced initially.

\begin{figure}[h]
\begin{center}
\fcolorbox{white}{white}{
  \begin{picture}(206,142) (279,-9)	
    \SetWidth{1.0}
    \SetColor{Black}
    \Line[arrow,arrowpos=0.5,arrowlength=5,arrowwidth=2,arrowinset=0.2](280,102)(360,82)
    \Line[arrow,arrowpos=0.5,arrowlength=5,arrowwidth=2,arrowinset=0.2](360,82)(440,102)
    \Photon(360,82)(360,32){5}{5}
 \Text(345,112)[lb]{\Large{\Black{[DIS]}}}
    \Text(280,112)[lb]{\Large{\Black{$\nu_\mu$}}}
    \Text(430,112)[lb]{\Large{\Black{$\mu^{ - }$}}}
    \Text(370,62)[lb]{\Large{\Black{$W^{ + }$}}}
    \Line[arrow,arrowpos=0.5,arrowlength=6.591,arrowwidth=2.636,arrowinset=0.2,double,sep=4](280,22)(350,22)
    \Line[arrow,arrowpos=0.5,arrowlength=5,arrowwidth=2,arrowinset=0.2](360,22)(440,42)
    \Line[arrow,arrowpos=0.5,arrowlength=5,arrowwidth=2,arrowinset=0.2](360,22)(440,32)
    \Line[arrow,arrowpos=0.5,arrowlength=5,arrowwidth=2,arrowinset=0.2](360,22)(440,22)
    \Line[arrow,arrowpos=0.5,arrowlength=5,arrowwidth=2,arrowinset=0.2](360,22)(440,12)
    \Line[arrow,arrowpos=0.5,arrowlength=5,arrowwidth=2,arrowinset=0.2](360,22)(440,2)
    \Line[arrow,arrowpos=0.5,arrowlength=5,arrowwidth=2,arrowinset=0.2](360,22)(440,-8)
    \Line[arrow,arrowpos=0.5,arrowlength=5,arrowwidth=2,arrowinset=0.2](360,22)(440,52)
    \GOval(360,22)(10,10)(0){0.882}
    \Text(450,35)[lb]{\LARGE{\Black{$\}\;$}}\Large{\Black{$ n\, \pi^{ \pm,0 }$}}}
    \Text(450,5)[lb]{\Large{\Black{$X$}}}
    \Text(280,37)[lb]{\Large{\Black{$N$}}}
  \end{picture}
\hspace{1.8cm}
  \begin{picture}(186,162) (279,-9)
    \SetWidth{1.0}
    \SetColor{Black}
    \Line[arrow,arrowpos=0.5,arrowlength=5,arrowwidth=2,arrowinset=0.2](280,122)(360,102)
    \Line[arrow,arrowpos=0.5,arrowlength=5,arrowwidth=2,arrowinset=0.2](360,102)(440,122)
    \Photon(360,102)(360,62){5}{4}
 \Text(345,132)[lb]{\Large{\Black{[COH]}}}
    \Text(280,132)[lb]{\Large{\Black{$\nu_\mu$}}}
    \Text(430,132)[lb]{\Large{\Black{$\mu^{ - }$}}}
    \Text(370,82)[lb]{\Large{\Black{$W^{ + }$}}}
    \SetWidth{2.0}
    \Line[arrow,arrowpos=0.5,arrowlength=7.727,arrowwidth=3.091,arrowinset=0.2,double,sep=4](280,2)(360,2)
    \SetWidth{1.0}
    \Line[arrow,arrowpos=0.5,arrowlength=5,arrowwidth=2,arrowinset=0.2](360,52)(440,52)
    \Line[arrow,arrowpos=0.5,arrowlength=6.591,arrowwidth=2.636,arrowinset=0.2,double,sep=4](360,52)(360,2)
    \SetWidth{2.0}
    \Line[arrow,arrowpos=0.5,arrowlength=7.727,arrowwidth=3.091,arrowinset=0.2,double,sep=4](360,2)(440,2)
    \SetWidth{1.0}
    \GOval(360,2)(10,10)(0){0.882}
    \GOval(360,52)(10,10)(0){0.882}
    \Text(280,17)[lb]{\Large{\Black{$A$}}}
    \Text(430,17)[lb]{\Large{\Black{$A'$}}}
    \Text(430,62)[lb]{\Large{\Black{$\pi$}}}
  \end{picture}
}
\end{center}

\begin{center}
\fcolorbox{white}{white}{
  \begin{picture}(246,134) (269,-9)
    \SetWidth{1.0}
    \SetColor{Black}
    \Line[arrow,arrowpos=0.5,arrowlength=5,arrowwidth=2,arrowinset=0.2](280,94)(360,74)
    \Line[arrow,arrowpos=0.5,arrowlength=5,arrowwidth=2,arrowinset=0.2](360,74)(440,94)
    \Photon(360,74)(360,24){5}{5}
 \Text(345,104)[lb]{\Large{\Black{[RES]}}}
    \Line[arrow,arrowpos=0.5,arrowlength=5,arrowwidth=2,arrowinset=0.2](280,24)(360,24)
    \Line[arrow,arrowpos=0.5,arrowlength=5,arrowwidth=2,arrowinset=0.2](280,8)(360,9)
    \Line[arrow,arrowpos=0.5,arrowlength=5,arrowwidth=2,arrowinset=0.2](280,-6)(360,-6)
    \GOval(280,9)(15,10)(0){0.882}
    \Line[arrow,arrowpos=0.5,arrowlength=5,arrowwidth=2,arrowinset=0.2](360,24)(420,24)
    \Line[arrow,arrowpos=0.5,arrowlength=5,arrowwidth=2,arrowinset=0.2](360,-6)(420,-6)
    \Line[arrow,arrowpos=0.5,arrowlength=5,arrowwidth=2,arrowinset=0.2](360,9)(420,9)
    \GOval(360,9)(15,10)(0){0.882}
    \Line[arrow,arrowpos=0.5,arrowlength=5,arrowwidth=2,arrowinset=0.2](420,24)(470,24)
    \Line[arrow,arrowpos=0.5,arrowlength=5,arrowwidth=2,arrowinset=0.2](420,9)(470,9)
    \Line[arrow,arrowpos=0.5,arrowlength=5,arrowwidth=2,arrowinset=0.2](420,-6)(470,-6)
    \GOval(470,9)(15,10)(0){0.882}
    \Line[arrow,arrowpos=0.5,arrowlength=5,arrowwidth=2,arrowinset=0.2](420,24)(460,64)
    \Line[arrow,arrowpos=0.5,arrowlength=5,arrowwidth=2,arrowinset=0.2](423,19)(460,56)
    \GOval(420,9)(15,14)(0){0.882}
    \GOval(461,58)(14,10)(0){0.882}
    \Text(280,104)[lb]{\Large{\Black{$\nu_\mu$}}}
    \Text(430,104)[lb]{\Large{\Black{$\mu^{ - }$}}}
    \Text(370,54)[lb]{\Large{\Black{$W^{ + }$}}}
    \Text(278,5)[lb]{\Large{\Black{$p$}}}
    \Text(300,27)[lb]{\Black{$d$}}
    \Text(300,11)[lb]{\Black{$u$}}
    \Text(300,-4)[lb]{\Black{$u$}}
    \Text(390,27)[lb]{\Black{$u$}}
    \Text(390,11)[lb]{\Black{$u$}}
    \Text(390,-4)[lb]{\Black{$u$}}
    \Text(410,5)[lb]{\Black{$\Delta^{ ++ }$}}
    \Text(437,50)[lb]{\Black{$u$}}
    \Text(447,35)[lb]{\Black{$\bar{d}$}}
    \Text(452,27)[lb]{\Black{$d$}}
    \Text(452,11)[lb]{\Black{$u$}}
    \Text(452,-4)[lb]{\Black{$u$}}
    \Text(467,5)[lb]{\Large{\Black{$p$}}}
    \Text(457,54)[lb]{\Black{$\pi^{ + }$}}
  \end{picture}
}
\end{center}
\caption{The main types of charged current muon neutrino scattering on a free nucleon/nucleus that produce pions directly. From top left to bottom right are: Deep Inelastic Scattering (DIS), Coherent pion production, and Resonance production (RES). In the figure $N$ is a nucleon, $A$ is a nucleus and X represents the hadronic system excluding pions.}
\label{fig:reactions}
\end{figure}
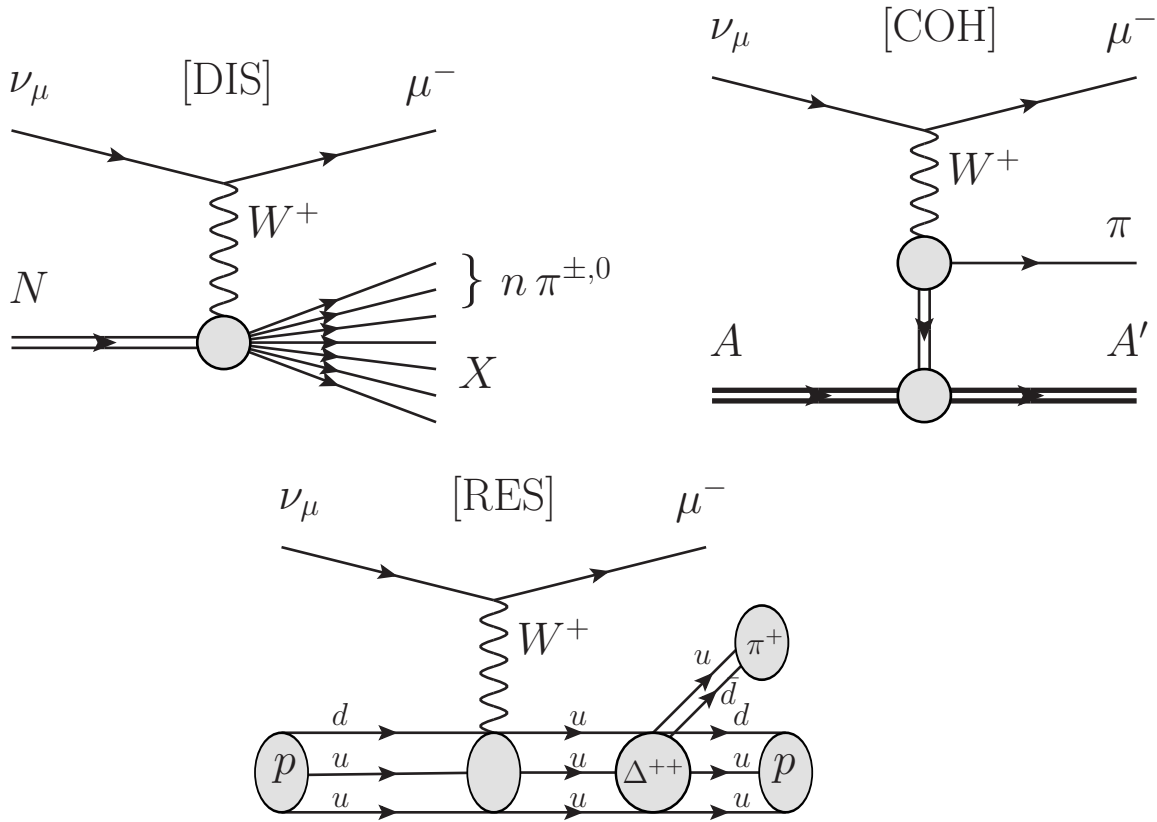

The dominant CC processes that produce pions directly are DIS,  COH and RES production. These are shown in Fig.~\ref{fig:reactions}.

As first proposed by Bodek and Ritchie~\cite{Bodek:1980ar} structure functions are used to describe DIS. Recently some progress in this field has been made using the ``higher twist'' QCD technique~\cite{Bodek:2002vp}. Neutrinos can also interact with the whole nucleus (instead of with individual nucleons as in the previous two processes) coherently producing pions. Typically COH pion production is described using the original Rein and Sehgal model~\cite{Rein:1983} with updates taking into account lepton mass terms~\cite{Rein:2006di}. A further description of COH scattering is presented in~\cite{AlvarezRuso:2007eg} and~\cite{Amaro:2008hd}. Resonant events are usually described using the Rein-Sehgal model~\cite{Rein:1980wg} describing the excitation of baryon resonances and pion production.

So far we have listed cross-section models which describe the scattering of neutrinos off free nucleons\footnote{With the notable exception of coherent pion production, which by its very nature is describing scattering off the whole nucleus.}. It is necessary to take into account the fact that these nucleons are not free but rather exist as bound states within a nuclear environment. The common approach within MC generators is to use the Relativistic Fermi gas (RFG) model where the Fermi motion of individual nucleons is taken into account. However its implementation often differs for different neutrino generators. In several papers \cite{Benhar:2006nr} the importance of considering Pauli blocking and FSI effects for $\nu_e + \nucl{16}{O} \to e X$ reactions (also applying to $\nu_\mu \nucl{16}{O}$) are shown. Also, it is shown by {\it O.~Benhar et al.} that the RFG model does not agree well with experimental data. A better description is offered through the use of spectral functions~\cite{Benhar:1994hw}, as measured in electron scattering experiments.

It is also necessary to describe hadronization, as well as the propagation of secondary particles out of the nucleus. The simulation must cover a description of both rescattering and absorption effects. A report on the modeling of final state interactions and the use of intranuclear cascade models  was presented at the school~\cite{DytmanLadek}. These are often individual features of a generator and are described in the next section.

\section{Results of Simulations for the Different Generators}
\label{sec:numbers}
For each of the generators we produced similar sets of 500,000 events simulating 1 GeV $\nu_{\mu}$ charged-current interactions on $\nucl{16}{O}$. We now present the results of these simulations in the form of tables showing the occupancy of primary and final state pion topologies. A primary state is defined as the topology of particles produced by the primary neutrino interaction and the final state is defined as the topology of the particles after any secondary interactions, such as intranuclear rescattering, have taken place. We do this separately for all of the generators. To aid fair comparison between the tables they are each preceded by a brief outline of the models and physics choices in each generator\footnote{We describe only those of relevance to charged current interactions.}.

\subsection{GENIE}
\label{sec:genie}
GENIE simulates neutrino interactions, for all neutrino flavors and all nuclear targets, over a large energy range from a few MeV to several hundred GeV. The physics models used can, broadly speaking, be split into models which describe cross-sections, hadronization, and nuclear physics. Full information on all the models and physics choices used in GENIE can be found at~\cite{Genie}. 

For cross-sections: Charged current quasi-elastic scattering is modeled using an implementation of the Llewellyn-Smith formalism using the latest BBA form factors~\cite{BBA_formfactors} as default. The production of baryon resonances, both neutral and charged current, is described using an implementation of the Rein-Sehgal model~\cite{Rein:1980wg} for which 16 baryon resonances are included\footnote{These are the 16 resonances, of the 18 listed in the original paper, listed as unambiguous in the latest PDG baryon tables~\cite{Amsler:2008zzb}.}. Coherent pion production is modeled using the Rein-Sehgal model~\cite{Rein:1983} with an updated PCAC formula from a recent revision~\cite{Rein:2006di} to the model that takes into account lepton mass terms. DIS interactions are calculated using an effective leading order model, with modifications at low $Q^{2}$ suggested by Bodek and Yang~\cite{Bodek:2003}. 

Hadronization is simulated using the AGKY model~\cite{AGKY}. It integrates an empirical low-invariant mass model with PYTHIA/JETSET at higher invariant mass and is tuned primarily to bubble chamber data on hydrogen and deuterium targets. There is a smooth transition between the models to ensure continuity of all simulated observables.

The effect of the nuclear environment is taken into account using an implementation of the Fermi Gas model with a modification by Bodek and Ritchie to include nucleon-nucleon correlations~\cite{Bodek:1981wr}. Other factors, such as Pauli blocking and the differences between nuclear and free nucleon structure functions are also taken into account. Intranuclear hadron transport is handled by the INTRANUKE/hA model. It is an effective, data-driven, model based on a wide range of hadron-nucleus and hadron-nucleon data. The model is validated through comparison to both pion and neutrino scattering data for nuclear targets.

The results of the simulation are shown in Table ~\ref{table:genie1}. For the simulation the GENIE default values for axial mass of $M_{A}^{QEL} = 0.99$ GeV, and  $M_{A}^{RES} = 1.12$ GeV, were used. The table shows the occupancy of primary and final state pion topologies. 

\begin{table*}[!h]
\centering
\scriptsize
\begin{tabular}{|p{1cm}||c|c|c|c|c|c|c|c|c|c|c||c|}
\hline
 & \multicolumn{11}{|c||}{Primary Hadronic System} & \\
\hline
Final State & $0\pi$ & $\pi^0$ & $\pi^+$ & $\pi^-$ & $2\pi^0$ & $2\pi^+$ & $2\pi^-$ & $\pi^0\pi^+$ & $\pi^0\pi^-$ & $\pi^+\pi^-$ & $\geq 3 \pi$ & Total \\
\hline\hline
$0\pi$ & 261866 & 9187 & 38142 & 0 & 21 & 54 & 0 & 161 & 0 & 54 & 1 & 309486 \\
\hline
$\pi^0$ & 0 & 32682 & 7085 & 0 & 127 & 14 & 0 & 546 & 0 & 22 & 2 & 40478 \\
\hline
$\pi^+$ & 549 & 890 & 139726 & 0 & 3 & 384 & 0 & 561 & 0 & 157 & 11 & 142281 \\
\hline
$\pi^-$ & 0 & 761 & 0 & 0 & 3 & 0 & 0 & 10 & 0 & 164 & 1 & 939 \\
\hline
$2\pi^0$ & 0 & 0 & 0 & 0 & 255 & 0 & 0 & 93 & 0 & 2 & 3 & 353 \\
\hline
$2\pi^+$ & 0 & 1 & 150 & 0 & 0 & 988 & 0 & 41 & 0 & 0 & 11 & 1191 \\
\hline
$2\pi^-$ & 0 & 0 & 0 & 0 & 1 & 0 & 0 & 0 & 0 & 0 & 0 & 1 \\
\hline
$\pi^0\pi^+$ & 542 & 194 & 610 & 0 & 10 & 58 & 0 & 2404 & 0 & 32 & 36 & 3886 \\
\hline
$\pi^0\pi^-$ & 0 & 0 & 0 & 0 & 8 & 0 & 0 & 2 & 0 & 36 & 0 & 46 \\
\hline
$\pi^+\pi^-$ & 237 & 0 & 0 & 0 & 0 & 0 & 0 & 48 & 0 & 594 & 33 & 912 \\
\hline
$\geq 3 \pi$ & 0 & 47 & 161 & 0 & 1 & 1 & 0 & 2 & 0 & 0 & 215 & 427 \\
\hline\hline
Total & 263194 & 43762 & 185874 & 0 & 429 & 1499 & 0 & 3868 & 0 & 1061 & 313
& 500000 \\
\hline
\end{tabular}\vspace{4pt}
\caption{Occupancy of primary and final state hadronic systems for a 500,000 event GENIE (release 2.5.1 was used) sample of $\nu_{\mu}$ on $\nucl{16}{O}$ for CC interactions only. A Fermi gas nuclear model and a default value of $M_{A}^{QEL} = 0.99$ GeV were used. The primary and final state systems were separated into different topological groups based on the number of pions.} 
\label{table:genie1}
\end{table*}

\subsection{NEUT}
\label{sec:neut}
NEUT is able to simulate neutrino interactions from 100 MeV up to a few TeV. The Rein-Sehgal model~\cite{Rein:1983},~\cite{Rein:1980wg} is used to simulate resonance and coherent pion production. The GRV94 and GRV98 pdfs~\cite{Gluck:1994uf} with Bodek-Yang corrections~\cite{Bodek:2004pc} are used to describe DIS events. Finally, QE events are simulated using the Llewellyn-Smith~\cite{Llewellyn Smith:1971zm} and Smith-Moniz~\cite{Kuzmin:2006dt} models. There are two recommended default values for the axial mass of 1.11 GeV or 1.21 GeV; these apply to both QE and RES interactions. For the simulations presented here  a value of 1.11 GeV was used for both. Nucleon rescattering and meson interactions (especially low momentum pions) in the nucleus are also modeled. To describe nuclear effects NEUT uses the cascade model. Each particle is tracked in the nucleus until it escapes. For low momentum pions ($p < 500$ MeV) mean free paths for absorption and inelastic scattering are calculated using the {\it L.L~Salcedo et al} model~\cite{Salcedo:1987md}. Higher momentum pion ($p>500$ MeV) parameters are taken from experimental results. Nucleon re-scattering is simulated considering elastic scattering and single/double pion production. 

The results of the simulation using the NEUT generator  are presented in Table~\ref{table:neut}. As before the table shows the various pion topologies before and after secondary interactions.

\begin{table*}[!h]
\centering
\scriptsize
\begin{tabular}{|p{1cm}||c|c|c|c|c|c|c|c|c|c|c||c|}
\hline
 & \multicolumn{11}{|c||}{Primary Hadronic System} & \\
\hline
Final State & $0\pi$ & $\pi^0$ & $\pi^+$ & $\pi^-$ & $2\pi^0$ & $2\pi^+$ &
$2\pi^-$ & $\pi^0\pi^+$ & $\pi^0\pi^-$ & $\pi^+\pi^-$ & $\geq 3 \pi$ & Total \\
\hline\hline
$0\pi$ & 328380 & 7148 & 33582 & 0 & 195 & 211 & 0 & 775 & 0 & 491 & 1 & 370783 \\
\hline
$\pi^0$ & 2224 & 14498 & 7119 & 0 & 616 & 75 & 0 & 1447 & 0 & 172 & 13 & 26164 \\
\hline
$\pi^+$ & 2899 & 1709 & 80376 & 0 & 60 & 702 & 0 & 1360 & 0 & 812 & 12 & 87930 \\
\hline
$\pi^-$ & 743 & 1632 & 742 & 0 & 66 & 6 & 0 & 126 & 0 & 899 & 9 & 4223 \\
\hline
$2\pi^0$ & 11 & 99 & 92 & 0 & 711 & 5 & 0 & 270 & 0 & 13 & 15 & 1216 \\
\hline
$2\pi^+$ & 2 & 9 & 313 & 0 & 4 & 872 & 0 & 245 & 0 & 16 & 7 & 1468 \\
\hline
$2\pi^-$ & 2 & 6 & 2 & 0 & 6 & 0 & 0 & 2 & 0 & 6 & 1 & 25 \\
\hline
$\pi^0\pi^+$ & 14 & 114 & 508 & 0 & 115 & 141 & 0 & 3395 & 0 & 150 & 27 & 4464 \\
\hline
$\pi^0\pi^-$ & 10 & 71 & 44 & 0 & 113 & 2 & 0 & 47 & 0 & 195 & 7 & 489 \\
\hline
$\pi^+\pi^-$ & 31 & 78 & 259 & 0 & 12 & 16 & 0 & 270 & 0 & 2195 & 21 &2882 \\
\hline
$\geq 3 \pi$ & 1 & 49 & 142 & 0 & 8 & 5 & 0 & 30 & 0 & 23 & 98 & 356 \\
\hline\hline
Total & 334317 & 25413 & 123179 & 0 & 1906 & 2035 & 0 & 7967 & 0 & 4972 & 211 & 500000 \\
\hline
\end{tabular}\vspace{4pt}
\caption{Occupancy of primary and final state hadronic systems for a 500,000 event NEUT sample of $\nu_{\mu}$ on $\nucl{16}{O}$ for CC interactions only. A Fermi gas nuclear model and a default value of $Ma^{QEL} = 1.11$ GeV were used. The primary and final state systems were separated into different topological groups based on the number of pions.} 
\label{table:neut}
\end{table*}

\subsection{FLUKA}
\label{sec:fluka}
FLUKA simulates the transport and interaction of particles with a focus on the description of nuclear models. Particle interactions are described using a Generalized IntraNuclear Cascade (GINC)~\cite{Fluka}. FLUKA now also describes neutrino interactions. QE processes have been included since 1997 and recently the capability to simulate DIS and RES neutrino interactions has been added using the NunDIS and NunRES generators. Hadron-nucleon interactions and nuclear effects are taken into account using the main reaction mechanism, called PEANUT~\cite{Fluka}. For example, pion-nucleon interactions can proceed through the non-resonant and the p-wave resonant channel, with the formation of a $\Delta$ resonance. The resonance can then either interact with other nucleons or decay. These effects can lead to pion absorption or to elastic scattering and charge exchange.

We present the results of the simulation for FLUKA in Table~\ref{table:fluka}.  It is important to mention that the recent DIS generator in FLUKA is a beta version and some of the pion production numbers in Table~\ref{table:fluka} are expected to be quite high. Also there is a difference between this sample and those produced using the other generators because at present FLUKA does not simulate COH pion production and so this process is left out. 

\begin{table*}[!h]
\centering
\scriptsize
\begin{tabular}{|p{1cm}||c|c|c|c|c|c|c|c|c|c|c||c|}
\hline
 & \multicolumn{11}{|c||}{Primary Hadronic System} & \\
\hline
Final State & $0\pi$ & $\pi^0$ & $\pi^+$ & $\pi^-$ & $2\pi^0$ & $2\pi^+$ &
$2\pi^-$ & $\pi^0\pi^+$ & $\pi^0\pi^-$ & $\pi^+\pi^-$ & $\geq 3 \pi$ & Total \\
\hline\hline
$0\pi$ & 267108 & 13702 & 41299 & 0 & 47 & 4 & 0 & 129 & 0 & 25 & 0 & 322314 \\
\hline
$\pi^0$ & 964 & 38871 & 8453 & 0 & 307 & 4 & 0 & 465 & 0 & 9 & 0 & 49073 \\
\hline
$\pi^+$ & 2005 & 2903 & 116159 & 0 & 16 & 29 & 0 & 421 & 0 & 91 & 2 & 121626 \\
\hline
$\pi^-$ & 121 & 1989 & 438 & 0 & 18 & 0 & 0 & 26 & 0 & 89 & 0 & 2681 \\
\hline
$2\pi^0$ & 0 & 115 & 122 & 0 & 548 & 1 & 0 & 68 & 0 & 0 & 0 & 854 \\
\hline
$2\pi^+$ & 2 & 11 & 230 & 0 & 2 & 37 & 0 & 61 & 0 & 0 & 2 & 345 \\
\hline
$2\pi^-$ & 0 & 1 & 1 & 0 & 0 & 0 & 0 & 0 & 0 & 0 & 0 & 2 \\
\hline
$\pi^0\pi^+$ & 2 & 177 & 391 & 0 & 53 & 6 & 0 & 1383 & 0 & 21 & 0 & 2033 \\
\hline
$\pi^0\pi^-$ & 0 & 47 & 15 & 0 & 38 & 0 & 0 & 9 & 0 & 14 & 0 & 124 \\
\hline
$\pi^+\pi^-$ & 0 & 118 & 320 & 0 & 1 & 0 & 0 & 78 & 0 & 392 & 1 & 910 \\
\hline
$\geq 3 \pi$ & 0 & 7 & 13 & 0 & 1 & 0 & 0 & 5 & 0 & 0 & 12 & 38 \\
\hline\hline
Total & 270202 & 57941 & 167441 & 0 & 1031 & 81 & 0 & 2645 & 0 & 641 & 18 & 500000 \\
\hline
\end{tabular}\vspace{4pt}
\caption{Occupancy of primary and final state hadronic systems for a 500,000 event FLUKA sample of $\nu_{\mu}$ on $\nucl{16}{O}$ for CC interactions only. The primary and final state systems were separated into different topological groups based on the number of pions.} 
\label{table:fluka}
\end{table*}
 
\subsection{NuWro}
\label{sec:nuwro}
NuWro uses the Llewellyn Smith model~\cite{Llewellyn Smith:1971zm} with the latest BBA form factors~\cite{BBA_formfactors} for QES events. DIS processes are described using the GRV94 pdf with Bodek-Yang corrections. The Rein-Sehgal models~\cite{Rein:1983} and \cite{Rein:2006di} are used to describe COH pion production. We now give a more detailed description of how these processes are implemented in NuWro~\cite{Jan_communications} :
\begin{itemize}
 \item The RES region is defined by a cut at $W < 1.4$ GeV, where $W$ is the invariant hadronic mass and has the standard definition\footnote{$W^{ 2 } = (P+q)^{ 2 }$, where $P$ is initial nucleon 4-momentum and $- q^{ 2 } = Q^{ 2 }$ is the squared 4-momentum transfer.}. Only $\Delta$ resonance is considered with form factors from a fit to ANL and BNL single pion production data~\cite{Graczyk}. The non-resonant background is modelled as a fraction of the DIS contribution.
 \item In the region of $1.4 \ \mathrm{GeV} < W < 1.6 \ \mathrm{GeV}$, the $\Delta$ contribution is linearly turned off as the DIS contribution is turned on.
 \item The DIS models are applied when $W > 1.6$ GeV. Together with the Bodek-Yang model NuWro's own hadronization model is used.
 \item The COH pion production contribution, as predicted by the Rein-Sehgal model, is multiplied by a factor of 0.66.
\end{itemize}

NuWro offers a choice of two basic nuclear models: a relativistic fermi gas (RFG) model or an effective spectral function model. For the tables shown here the RFG model was used. We also used the default value for the axial mass of $M_A = $ 1.1 GeV. In the appendix Tables~\ref{table:nuwro_fermigas_1} and \ref{table:nuwro_spectralfunction1000} show results for samples generated with a non-default axial mass and for both the RFG and effective spectral function nuclear models. 

\subsubsection{NuWro's Own Intranuclear Cascade Model}
\label{sec:NuwroAndOwnHadronTran}
For this report two different intranuclear cascades were used, one included with NuWro and another, the Bertini cascade, that comes from Geant4. First we show the results using NuWro's intranuclear cascade in Table~\ref{table:nuwro_geantcomparison1}.  

\begin{table*}[!h]
\centering
\scriptsize
\begin{tabular}{|p{1cm}||c|c|c|c|c|c|c|c|c|c|c||c|}
\hline
 & \multicolumn{11}{|c||}{Primary Hadronic System} & \\
\hline
Final State & $0\pi$ & $\pi^0$ & $\pi^+$ & $\pi^-$ & $2\pi^0$ & $2\pi^+$ & $2\pi^-$ & $\pi^0\pi^+$ & $\pi^0\pi^-$ & $\pi^+\pi^-$ & $\geq 3 \pi$ & Total \\
\hline\hline
$0\pi$ & 281781 & 10437 & 49381 & 1 & 49 & 78 & 0 & 523 & 0 & 223 & 2 & 342575 \\ 
\hline
$\pi^0$ & 2137 & 17826 & 12404 & 0 & 178 & 29 & 0 & 1095 & 0 & 85 & 24 & 33778 \\ 
\hline
$\pi^+$ & 5241 & 3168 & 97356 & 0 & 32 & 255 & 0 & 1187 & 0 & 407 & 4 & 107650 \\ 
\hline
$\pi^-$ & 1165 & 2849 & 1667 & 1 & 22 & 4 & 0 & 140 & 0 & 413 & 4 & 6265 \\ 
\hline
$2\pi^0$ & 12 & 332 & 123 & 0 & 179 & 4 & 0 & 203 & 0 & 13 & 38 & 904 \\ 
\hline
$2\pi^+$ & 11 & 39 & 1948 & 0 & 3 & 248 & 0 & 245 & 0 & 10 & 4 & 2508 \\ 
\hline
$2\pi^-$ & 0 & 20 & 2 & 0 & 1 & 0 & 0 & 1 & 0 & 11 & 2 & 37 \\ 
\hline
$\pi^0\pi^+$ & 26 & 382 & 1090 & 0 & 38 & 65 & 0 & 1963 & 0 & 100 & 18 & 3682 \\ 
\hline
$\pi^0\pi^-$ & 7 & 237 & 94 & 0 & 37 & 0 & 0 & 55 & 0 & 76 & 7 & 513 \\ 
\hline
$\pi^+\pi^-$ & 22 & 67 & 783 & 0 & 6 & 8 & 0 & 220 & 0 & 804 & 8 & 1918 \\ 
\hline
$\geq 3 \pi$ & 1 & 27 & 47 & 0 & 10 & 11 & 0 & 69 & 0 & 29 & 76 & 270 \\ 
\hline\hline
Total & 290403 & 35384 & 164895 & 2 & 555 & 702 & 0 & 5701 & 0 & 2171 & 187 & 500000 \\ 
\hline
\end{tabular}\vspace{4pt}
\caption{Occupancy of primary and final state hadronic systems for a 500,000 event NuWro sample of $\nu_{\mu}$ on $\nucl{16}{O}$ for CC interactions only. Sample generated with NuWro v112. $M_{A}^{ QEL_{CC} }$ = 1.10 GeV, nuclear model: Fermi gas. The primary and final state systems were separated into different topological groups based on the number of pions.} 
\label{table:nuwro_geantcomparison1}
\end{table*}

\subsubsection{Combination of NuWro and Geant4 generators}
\label{sec:NuwroAndGeant4}
Now we show the results for the simulation using the Geant4~\cite{Geant4} intranuclear cascade. The primary neutrino interactions on oxygen are generated in NuWro and secondary particles are propagated inside nucleus using the Bertini cascade~\cite{bertini_cascade}.

There are two intranuclear cascades in Geant4: Bertini and Binary. However we only consider the Bertini cascade. It works very well for particles with energy below 3 GeV. It takes into account a variety of interactions that nucleons and pions can undergo, using very recent cross-sections. Compared with Binary, the Bertini cascade is better tested and developed by both members and users of Geant4. A detailed description can be found in~\cite{bertini_cascade}. 

We would like to state that, in its original form, the Bertini cascade was not set up to propagate secondary particles from neutrino interactions. A particle from a neutrino interaction can originate at any point within the volume of the nucleus. This meant changing the original Bertini cascade, in which a particle always hits a nucleus from the outside (so that the cascade always starts on the surface of the nucleus) so that the point where the particle enters is selected uniformly over the volume of the nucleus. The results of the simulation using this approach are shown in Table~\ref{table:nuwro_geantcomparison2}.

\begin{table*}[!h]
\centering
\scriptsize
\begin{tabular}{|p{1cm}||c|c|c|c|c|c|c|c|c|c|c||c|}
\hline
 & \multicolumn{11}{|c||}{Primary Hadronic System} & \\
\hline
Final State & $0\pi$ & $\pi^0$ & $\pi^+$ & $\pi^-$ & $2\pi^0$ & $2\pi^+$ & $2\pi^-$ & $\pi^0\pi^+$ & $\pi^0\pi^-$ & $\pi^+\pi^-$ & $\geq 3 \pi$ & Total \\
\hline\hline
$0\pi$ & 282443 & 8888 & 36738 & 0 & 67 & 47 & 0 & 616 & 0 & 173 & 18 & 328990 \\ 
\hline
$\pi^0$ & 3148 & 20328 & 8963 & 0 & 147 & 25 & 0 & 801 & 0 & 65 & 25 & 33502 \\ 
\hline
$\pi^+$ & 3866 & 2770 & 115609 & 0 & 31 & 218 & 0 & 1162 & 0 & 323 & 6 & 123985 \\ 
\hline
$\pi^-$ & 908 & 2845 & 1190 & 2 & 34 & 1 & 0 & 134 & 0 & 430 & 10 & 5554 \\ 
\hline
$2\pi^0$ & 8 & 123 & 152 & 0 & 208 & 1 & 0 & 91 & 0 & 4 & 24 & 611 \\ 
\hline
$2\pi^+$ & 7 & 32 & 676 & 0 & 1 & 363 & 0 & 199 & 0 & 1 & 1 & 1280 \\ 
\hline
$2\pi^-$ & 0 & 8 & 3 & 0 & 4 & 0 & 0 & 3 & 0 & 9 & 0 & 27 \\ 
\hline
$\pi^0\pi^+$ & 12 & 161 & 904 & 0 & 30 & 41 & 0 & 2448 & 0 & 53 & 10 & 3659 \\ 
\hline
$\pi^0\pi^-$ & 2 & 89 & 70 & 0 & 26 & 0 & 0 & 30 & 0 & 49 & 9 & 275 \\ 
\hline
$\pi^+\pi^-$ & 9 & 135 & 578 & 0 & 5 & 4 & 0 & 197 & 0 & 1056 & 22 & 2006 \\ 
\hline
$\geq 3 \pi$ & 0 & 5 & 12 & 0 & 2 & 2 & 0 & 20 & 0 & 8 & 62 & 111 \\ 
\hline\hline
Total & 290403 & 35384 & 164895 & 2 & 555 & 702 & 0 & 5701 & 0 & 2171 & 187 & 500000 \\ 
\hline
\end{tabular}\vspace{4pt}
\caption{Occupancy of primary and final state hadronic systems for a 500,000 event NuWro (plus GEANT4) sample of $\nu_{\mu}$ on $\nucl{16}{O}$ for CC interactions only. Sample generated with NuWro v112. $M_{A}^{ QEL_{CC} }$ = 1.10 GeV, nuclear model: Fermi gas. Also, hadrons produced inside the nucleus were propagated using the  GEANT4 cascade model. The primary and final state systems were separated into different topological groups based on the number of pions.} 
\label{table:nuwro_geantcomparison2}
\end{table*}

\newpage
\section{Discussion and Conclusions}
\label{sec:conclusions}

Looking at the tables for each generator we can see some large differences in both the primary and final state topologies (number of pions at the initial and final states). In many cases these differences are above statistical fluctuations\footnote{In some of the cells with enough high statistics these differences can be above 10\%, well above statistical fluctuations. } and reflect the difficulty in modeling this energy region. At $\sim 1$ GeV we are in a transition region where QE, RES processes dominate but where there is also a significant DIS component being switched on as we increase in energy. Although the models used to describe these processes separately are often common to a generator, there are still many differences in the way in which a particular generator handles the merging of the relative contributions in this transition region. This combined with differences in the assumed nominal values for many of the input parameters can lead to different predictions from the same set of models. 

In general QE processes give rise to topologies with no pions in the initial and final states whereas DIS and RES are more likely to result in events with pions in the primary and final state. All the generators have a larger number of $0\pi$ topologies in the final state than were in the primary state. This indicates that pions are more likely to be absorbed than created. One noticeable feature is that NEUT has a larger number of $0\pi$ topologies in both the initial and final states than the other generators. This is an indicator that either the cross-section for QE processes in NEUT is higher than in the other generators or that the contribution from DIS and RES events is lower. Mainly we expect differences to depend on the values of form--factor parameters and the nuclear models used in each generator\footnote{Or, if available, the choice of simulation within a particular generator.}. 

Resonant events are tricky to handle; there are large uncertainties on the underlying cross-sections as well as differences in the way each generator models the propagation of pions in the nuclear environment. Which model is used is important and we can see the effect of these differences in the tables presented in this report. For instance, if we look at the $0\pi$ primary state column we see that the number of these events which result in a single $\pi^{+}$ in the final state topology varies from 549 (Table~\ref{table:genie1}) all the way up to 5241 (Table~\ref{table:nuwro_geantcomparison1}). Still looking at the first column we see that the number of $\pi^{0}$'s and $\pi^{-}$'s in the final state can vary from zero (Table~\ref{table:genie1}) to 1--3 thousands (Tables~\ref{table:neut},~\ref{table:fluka},~\ref{table:nuwro_geantcomparison1},~\ref{table:nuwro_geantcomparison2}). All of these events had no pions in the primary state and so these numbers directly reflect differences in the hadron transport models used in the generators.

As mentioned before the tables presented so far contain important information about FSIs. To elucidate the effects of FSIs we have compiled a summary table (Table~\ref{table:fsieffect}) showing directly the topology changing effect due to intranuclear hadron transport. For each of the generators, and for a selection of primary and final state topologies, we show, out of all events with a given primary state topology, the fraction of those which have both the primary and final state topology. Looking at Table~\ref{table:fsieffect} the first two rows tell us the fraction of events, with a single pion in the initial state, that will still have a single pion in the final state. We can see that the most probable scenario is that a pion created at the primary vertex will not re-interact (we can look at this as a transparency of the nucleus). The next two rows show the percent of pions which are absorbed and the remaining rows show the effect  due to charge exchange processes (we assume that a pion, to first order, re-scatters only once). It seems that NuWro has a lower transparency compared to the other generators, whilst GENIE's is higher than that of the others. This may be in response to the absorption and charge exchange processes, for which GENIE could have too little and NuWro too much. Despite these differences the agreement is still good, considering the complexity of FSIs, and one can convince oneself that the analyzed MCs give quite similar results. These modes are important to neutrino oscillation experiments since single pion events form the main background channel.

\begin{table*}[!h]
\centering
\small
\renewcommand{\arraystretch}{1.1}
\begin{tabular}{|c|c|c|c|c|c|c|}
\hline
 & GENIE & NEUT & FLUKA & NuWro & NuWro + G4 \\
\hline
\hline  $\pi^0\rightarrow\pi^0$ & 75\% & 57\% & 67\% & 50\% & 57\% \\ 
\hline  $\pi^+\rightarrow\pi^+$ & 75\% & 65\% & 69\% & 59\% & 70\% \\ 
\hline  $\pi^0\rightarrow 0\pi's$ & 20\% & 28\% & 24\% & 29\% & 25\% \\ 
\hline  $\pi^+\rightarrow 0\pi's$ & 20\% & 27\% & 25\% & 30\% & 22\% \\ 
\hline  $\pi^0\rightarrow\pi^+$ & 2\% & 7\% & 5\% & 9\% & 8\% \\ 
\hline  $\pi^0\rightarrow\pi^-$ & 2\% & 6\% & 3\% & 8\% & 8\% \\ 
\hline  $\pi^+\rightarrow\pi^0$ & 4\% & 6\% & 5\% & 8\% & 5\% \\ 
\hline 
\end{tabular}\vspace{4pt}
\caption{Rate of events with single pion or no pion in final state if there was single pion in initial state.}
\label{table:fsieffect}
\end{table*}

Taking into accout recent MiniBooNE results~\cite{miniboone} of CC$1 \pi^{ + }$ to CCQE cross section ratios we can approximately compare the corresponding numbers from our simulations with each other and with the MiniBooNE measurements. We present this comparison in Table~\ref{table:MiniBooNE} in the appendix.

In this report we have presented event numbers, with an emphasis on pion production, for a large number of the current neutrino generators. We have started to discuss the differences between the different generators and models. These differences are now a matter for further investigation. A more extensive set of tables has been published on the Ladek MC Project website and more tables, at different energies, are planned to be produced in the near future. We hope that the tables will be a valuable resource to the reader. Improvements in the theoretical models used by generators, and better measurements made by future neutrino experiments, will, by improving our understanding of neutrino cross-sections, help shed light on the elusive nature of the neutrino.

\begin{ack}
We would like to acknowledge the organizers and lecturers of the Ladek Winter School, especially Jan Sobczyk and Costas Andreopoulos for the support and helpful comments.
\end{ack}

\newpage
\section{Appendix}
\label{NuWro:appendix}	
In this appendix we present some tables showing the effect of changing the values for the axial mass parameter or the use of a different nuclear model. We also present some results from the Nuance generator which, due to problems with generating samples comparable to those produced using the other generators, were left out of the main report. Finally we show some comparisons to a MiniBooNE measurement of the CC single pion-production to quasi-elastic cross-section ratio.


{\bf GENIE}\\

\begin{table*}[!h]
\centering
\scriptsize
\begin{tabular}{|p{1cm}||c|c|c|c|c|c|c|c|c|c|c||c|}
\hline
 & \multicolumn{11}{|c||}{Primary Hadronic System} & \\
\hline
Final State & $0\pi$ & $\pi^0$ & $\pi^+$ & $\pi^-$ & $2\pi^0$ & $2\pi^+$ & $2\pi^-$ & $\pi^0\pi^+$ & $\pi^0\pi^-$ & $\pi^+\pi^-$ & $\geq 3 \pi$ & Total \\
\hline\hline
$0\pi$ & 282947 & 8263 & 34477 & 0 & 18 & 46 & 0 & 134 & 0 & 39 & 1 & 325925 \\ 
\hline
$\pi^0$ & 0 & 29903 & 6448 & 0 & 120 & 24 & 0 & 489 & 0 & 12 & 6 & 37002 \\ 
\hline
$\pi^+$ & 3930 & 768 & 123912 & 0 & 3 & 368 & 0 & 531 & 0 & 140 & 13 & 129665 \\ 
\hline
$\pi^-$ & 0 & 710 & 0 & 0 & 2 & 0 & 0 & 15 & 0 & 151 & 4 & 882 \\ 
\hline
$2\pi^0$ & 0 & 0 & 0 & 0 & 223 & 4 & 0 & 84 & 0 & 1 & 8 & 320 \\ 
\hline
$2\pi^+$ & 0 & 1 & 144 & 0 & 0 & 946 & 0 & 53 & 0 & 0 & 17 & 1161 \\ 
\hline
$2\pi^-$ & 0 & 0 & 0 & 0 & 0 & 0 & 0 & 0 & 0 & 0 & 0 & 0 \\ 
\hline
$\pi^0\pi^+$ & 646 & 192 & 521 & 0 & 4 & 67 & 0 & 2233 & 0 & 24 & 37 & 3724 \\ 
\hline
$\pi^0\pi^-$ & 0 & 0 & 0 & 0 & 7 & 0 & 0 & 2 & 0 & 38 & 3 & 50 \\ 
\hline
$\pi^+\pi^-$ & 218 & 1 & 0 & 0 & 1 & 0 & 0 & 44 & 0 & 583 & 22 & 869 \\ 
\hline
$\geq 3 \pi$ & 0 & 31 & 161 & 0 & 1 & 1 & 0 & 6 & 0 & 1 & 201 & 402 \\ 
\hline\hline
Total & 287741 & 39869 & 165663 & 0 & 379 & 1456 & 0 & 3591 & 0 & 989 & 312 & 500000 \\ 
\hline
\end{tabular}\vspace{4pt}
\caption{Occupancy of primary and final state hadronic systems for a 500,000 event GENIE sample of $\nu_{\mu}$ on $\nucl{16}{O}$ for CC interactions only. A Fermi gas nuclear model and a non default value of MaQEL = 1.18 GeV were used. The primary and final state systems were separated into different topological groups based on the number of pions. Coherent pion production events were counted as having a single pion in the final state only.} 
\label{table:genie2}
\end{table*}


{\bf NuWro}\\

There are two additional tables for NuWro event samples. In Table~\ref{table:nuwro_fermigas_1} a Fermi gas nuclear model was used but with an axial mass equal to 1.0 GeV. In Table~\ref{table:nuwro_spectralfunction1000} an effective spectral function was used instead of the Fermi gas model.

\begin{table*}[!h]
\centering
\scriptsize
\begin{tabular}{|p{1cm}||c|c|c|c|c|c|c|c|c|c|c||c|}
\hline
 & \multicolumn{11}{|c||}{Primary Hadronic System} & \\
\hline
Final State & $0\pi$ & $\pi^0$ & $\pi^+$ & $\pi^-$ & $2\pi^0$ & $2\pi^+$ & $2\pi^-$ & $\pi^0\pi^+$ & $\pi^0\pi^-$ & $\pi^+\pi^-$ & $\geq 3 \pi$ & Total \\
\hline\hline
$0\pi$ & 266733 & 11177 & 52296 & 0 & 49 & 51 & 0 & 547 & 0 & 237 & 1 & 331091 \\ 
\hline
$\pi^0$ & 1876 & 19692 & 12988 & 1 & 198 & 26 & 0 & 1188 & 0 & 89 & 16 & 36074 \\ 
\hline
$\pi^+$ & 4701 & 3186 & 106161 & 0 & 20 & 276 & 0 & 1231 & 0 & 455 & 4 & 116034 \\ 
\hline
$\pi^-$ & 1085 & 2901 & 1641 & 0 & 19 & 2 & 0 & 124 & 0 & 449 & 7 & 6228 \\ 
\hline
$2\pi^0$ & 13 & 345 & 110 & 0 & 210 & 5 & 0 & 251 & 0 & 7 & 53 & 994 \\ 
\hline
$2\pi^+$ & 15 & 42 & 2116 & 0 & 1 & 288 & 0 & 237 & 0 & 10 & 0 & 2709 \\ 
\hline
$2\pi^-$ & 3 & 14 & 9 & 0 & 2 & 0 & 0 & 3 & 0 & 14 & 0 & 45 \\ 
\hline
$\pi^0\pi^+$ & 22 & 426 & 1180 & 0 & 37 & 62 & 0 & 2093 & 0 & 96 & 16 & 3932 \\ 
\hline
$\pi^0\pi^-$ & 3 & 241 & 94 & 0 & 37 & 1 & 0 & 48 & 0 & 96 & 14 & 534 \\ 
\hline
$\pi^+\pi^-$ & 27 & 61 & 845 & 0 & 1 & 14 & 0 & 220 & 0 & 864 & 14 & 2046 \\ 
\hline
$\geq 3 \pi$ & 0 & 15 & 66 & 0 & 13 & 14 & 0 & 97 & 0 & 27 & 81 & 313 \\ 
\hline\hline
Total & 274478 & 38100 & 177506 & 1 & 587 & 739 & 0 & 6039 & 0 & 2344 & 206 & 500000 \\ 
\hline
\end{tabular}\vspace{4pt}
\caption{Occupancy of primary and final state hadronic systems for a 500,000 event NuWro sample of $\nu_{\mu}$ on $\nucl{16}{O}$ for CC interactions only. Sample generated with NuWro v112. $M_{A}^{ QEL_{CC} }$ = 1.0 GeV, nuclear model: Fermi gas. The primary and final state systems were separated into different topological groups based on the number of pions.} 
\label{table:nuwro_fermigas_1}
\end{table*}

\begin{table*}[!h]
\centering
\scriptsize
\begin{tabular}{||p{1cm}||c|c|c|c|c|c|c|c|c|c|c||c||}
\hline
 & \multicolumn{11}{|c||}{Primary Hadronic System} & \\
\hline
Final State & $0\pi$ & $\pi^0$ & $\pi^+$ & $\pi^-$ & $2\pi^0$ & $2\pi^+$ & $2\pi^-$ & $\pi^0\pi^+$ & $\pi^0\pi^-$ & $\pi^+\pi^-$ & $\geq 3 \pi$ & Total \\
\hline\hline
$0\pi$ & 278271 & 10535 & 50092 & 0 & 46 & 67 & 0 & 493 & 0 & 205 & 2 & 339711 \\ 
\hline
$\pi^0$ & 1960 & 18390 & 12291 & 0 & 170 & 19 & 0 & 1000 & 0 & 78 & 12 & 33920 \\ 
\hline
$\pi^+$ & 4787 & 3016 & 101433 & 0 & 13 & 245 & 0 & 1123 & 0 & 356 & 3 & 110976 \\ 
\hline
$\pi^-$ & 1125 & 2807 & 1631 & 0 & 12 & 1 & 0 & 90 & 0 & 399 & 3 & 6068 \\ 
\hline
$2\pi^0$ & 9 & 271 & 100 & 0 & 167 & 0 & 0 & 208 & 0 & 7 & 24 & 786 \\ 
\hline
$2\pi^+$ & 8 & 39 & 1987 & 0 & 6 & 259 & 0 & 178 & 0 & 11 & 1 & 2489 \\ 
\hline
$2\pi^-$ & 1 & 25 & 5 & 0 & 4 & 0 & 0 & 0 & 0 & 8 & 0 & 43 \\ 
\hline
$\pi^0\pi^+$ & 19 & 396 & 1046 & 0 & 46 & 51 & 0 & 1795 & 0 & 76 & 15 & 3444 \\ 
\hline
$\pi^0\pi^-$ & 5 & 242 & 61 & 0 & 27 & 0 & 0 & 27 & 0 & 70 & 5 & 437 \\ 
\hline
$\pi^+\pi^-$ & 27 & 53 & 791 & 0 & 5 & 7 & 0 & 209 & 0 & 804 & 7 & 1903 \\ 
\hline
$\geq 3 \pi$ & 1 & 25 & 45 & 0 & 5 & 9 & 0 & 65 & 0 & 23 & 50 & 223 \\ 
\hline\hline
Total & 286213 & 35799 & 169482 & 0 & 501 & 658 & 0 & 5188 & 0 & 2037 & 122 & 500000 \\ 
\hline
\end{tabular}\vspace{4pt}
\caption{Occupancy of primary and final state hadronic systems for a 500,000 event NuWro sample of $\nu_{\mu}$ on $\nucl{16}{O}$ for CC interactions only. Sample generated with NuWro v112. $M_{A}^{ QEL_{CC} }$ = 1.0 GeV, nuclear model: effective spectral function. The primary and final state systems were separated into different topological groups based on the number of pions.}
\label{table:nuwro_spectralfunction1000}
\end{table*}


{\bf Nuance}\\

Nuance is an advanced and freely available neutrino generator written by Dave Casper of the University of California (the version used in this study is 3.006). The program is most suitable for generating events on oxygen/water, as its FSI model was thoroughly tested for this target. Resonant and coherent/diffractive interactions are simulated using the Rein-Sehgal model. More details can be found in~\cite{Casper:2002sd}. Unfortunately in the case of Nuance it was not possible to create a single sample consisting of events with final state particles before and after the FSIs have been applied. Therefore we were unable to produce tables like those presented for the other generators. Instead, in Table~\ref{table:nuance} we present pion statistics for two separate samples independently, where one had FSIs turned off and in the other they were left on.

\begin{table*}[!h]
\centering
\scriptsize
\begin{tabular}{|p{1cm}||c|c|}
\hline
CC only & with FSI & without FSI \\
\hline\hline
$0\pi$ & 337431 & 292705 \\ 
\hline
$\pi^0$ & 35033 & 35532 \\ 
\hline
$\pi^+$ & 121242 & 169365 \\ 
\hline
$\pi^-$ & 3737 & 0 \\ 
\hline
$2\pi^0$ & 390 & 630 \\ 
\hline
$2\pi^+$ & 214 & 100 \\ 
\hline
$2\pi^-$ & 8 & 0 \\ 
\hline
$\pi^0\pi^+$ & 801 & 774 \\ 
\hline
$\pi^0\pi^-$ & 180 & 0 \\ 
\hline
$\pi^+\pi^-$ & 817 & 892 \\ 
\hline
$\geq 3 \pi$ & 147 & 2 \\ 
\hline
\end{tabular}\vspace{4pt}
\caption{Two samples, for muon neutrino interactions on oxygen, of 500,000 events each were used, one with FSIs turned on and one with FSIs turned off. The muon neutrinos had an initial energy of 1 GeV and only CC interactions were taken into account. A Fermi gas nuclear model was used. The default axial masses used were $M_{A} = 1.1 \ \mathrm{GeV}$ for resonant single pion production, $M_{A} = 1.3 \ \mathrm{GeV}$ for resonant multi-pion production and $M_{A} = 1.03 \ \mathrm{GeV}$ for coherent/diffractive pion production (the same value was used in QE event generation; the vector mass for QE was $M_{V} = 0.84 \ \mathrm{GeV}$). The samples consist of QE(59\%), RES(39\%) and COH(2\%) events.} 
\label{table:nuance}
\end{table*}

\newpage
{\bf The single charged pion production to quasi-elastic cross section ratios}\\

In Table~\ref{table:MiniBooNE} the comparison between the ratio of $1 \pi^{ + } / 0 \pi$ is shown for the generators we considered. From the MiniBooNE data~\cite{miniboone} re-scaled for an isoscalar target and corrected for FSI's (meaning events at the initial vertex and before any hadronic re-interactions) the following values  of the ratio are\footnote{The MiniBooNE ratios of CC1$\pi^{ + }$/CCQE are taken from Fig. 2~\cite{miniboone}.}: $0.52\pm0.06$ for a neutrino energy of $0.95\pm0.05$ GeV and $0.63\pm0.07$ for a neutrino energy of $1.05\pm0.05$ GeV. These numbers agree well with the corresponding numbers, see the {\it PHS total} column in Table~\ref{table:MiniBooNE}, obtained in our simulations, even though the Genie ratio is slightly higher and the Neut ratio is smaller.

\begin{table*}[!h]
\centering
\scriptsize
\begin{tabular}{|l|c|c|}
\hline
\multicolumn{3}{|c|}{Our simulations, Charged Current $\nu_\mu$ interactions on $\nucl{16}{O}$}	\\
\hline
	$E_\nu = 1$ GeV	&	$1 \pi^{ + } / 0 \pi$	(PHS total)	&	$1 \pi^{ + } / 0 \pi$ (Final State total)	\\
\hline\hline
	GENIE	&	0.706	&	0.460	\\
\hline
	NEUT	&	0.368	&	0.237	\\
\hline
	FLUKA	&	0.620	&	0.377	\\
\hline
	NuWro	&	0.568	&	0.314	\\
\hline
	NuWro+G4	&	0.568	&	0.377	\\
\hline
	Nuance	&	0.579	&	0.359	\\
\hline
\end{tabular}\vspace{4pt}
\caption{The ratios $1 \pi^{ + } / 0 \pi$ obtained in our simulations.}
\label{table:MiniBooNE}
\end{table*}

\end{document}